# Image simulation for biological microscopy: microlith

**Shalin B. Mehta[1,2,*] and Rudolf Oldenbourg[1,3]**

[1] *Cellular Dynamics Program, Marine Biological Laboratory, 7 MBL Street, Woods Hole, MA 02543, USA*
[3] *Physics Department, Brown University, Providence RI 02912, USA*
* [mshalin@mbl.edu, shalin.mehta@gmail.com](mailto:mshalin@mbl.edu)

**Abstract:** Image simulation remains under-exploited for the most widely used biological phase microscopy methods, because of difficulties in simulating partially coherent illumination. We describe an open-source toolbox, microlith (https://code.google.com/p/microlith), which accurately predicts three-dimensional images of a thin specimen observed with any partially coherent imaging system, including coherently illuminated and incoherent, self-luminous specimens. Its accuracy is demonstrated by comparing simulated and experimental bright-field and dark-field images of well-characterized amplitude and phase targets, respectively. The comparison provides new insights about the sensitivity of the dark-field microscope to mass distributions in isolated or periodic specimens at the length-scale of 10nm. Based on predictions using microlith, we propose a novel approach for detecting nanoscale structural changes in a beating axoneme using a dark-field microscope.

## 1. Introduction

Use of partially coherent (i.e., angularly diverse) illumination has been popular for biological imaging in the form of contrast mechanisms such as dark-field, phase-contrast, differential interference contrast, differential phase contrast, and polarized-light microscopy. Partially coherent illumination provides widely recognized benefits over spatially coherent illumination – mainly, higher resolution, improved immunity against optical imperfections in the light path, and a degree of depth sectioning. Among emerging partially coherent methods in optical and X-ray regimes, emphasis is on the quantitative measurement of the specimen's optical properties and their interpretation in terms of the underlying structure of the specimen. The optical properties of interest are absorption and phase or their polarization-resolved counterparts, diattenuation and birefringence, that report molecular order [1,2]. Quantitative analysis of the specimen and the design of an optimal microscope require accurate modeling of image formation. However, partially coherent illumination leads to bilinear image formation, i.e., the image intensity at a given point in image space depends not only on the corresponding point in the specimen, but on pairs of points in the specimen [3]. The intrinsic bilinearity has hampered the use of partially coherent imaging systems for quantitative biology despite the widely recognized experimental benefits. Accurate simulations can enable quantitative interpretation of partially coherent images, design of new systems, and design of new reconstruction algorithms.

Simulation and analysis of coherent and incoherent systems have been well established [4,5] thanks to their linear image formation. Linear image formation allows the description of an imaging system as a point spread function in space or a transfer function in spatial frequency for a general 3D specimen. Most of the partially coherent phase reconstruction and segmentation algorithms employ approximate linear models - either of a weak phase variation [6], slowly varying gradient [7–9], slowly varying curvature [10], or dictionary of diffraction patterns [11] – to retrieve a quantitative representation of the specimen phase. For partially coherent microscopes, the appropriate model for the specimen and the appropriate *transfer function* that relates the image with the optical/structural properties of the specimen depend on the problem being addressed and the contrast method. Therefore, flexible, problem-driven simulations are necessary for accurate interpretation.

While the optical lithography community regularly employs partially coherent simulation tools to keep pace with Moore's law, accurate simulation has remained under-exploited for biological, partially coherent microscopy. Apart from commercial image simulation tools such as PROLITH™, the lithography community has made available some free simulation tools (e.g., LAVA [12]), but they are not suitable for simulating biological contrast methods and none are open source.

With this paper and the associated open-source MATLAB toolbox, microlith [13], we aim to bridge this gap. The design goals for microlith are to allow quantitative comparison of biological microscopy methods, quantitative analysis of partially coherent images, and design of phase-retrieval approaches. microlith permits the flexibility of simulating 3D images of a thin specimen under any scalar partially coherent system. It allows simulation of spatially coherent imaging (when the illumination originates from a point) or incoherent imaging (when imaging fluorescent/self-luminous specimens) as special cases. Coherent image simulation algorithms in microlith have been used to simulate the 3D response of coherently illuminated phase filters designed to provide large depth of focus with high transverse resolution [14]. Incoherent image simulation algorithms implemented in microlith have been used to study the 3D imaging properties of a fluorescence super-resolution microscope that employs an area detector in a point-scanning confocal mode with pixel reassignment [15].

Apart from the optical microscopy community, the toolbox may also serve as an open source alternative for the optical lithography community - hence the name microlith. The project website provides a tutorial and a user guide along with simulations of partially coherent imaging problems for which analytical expressions are available and representative simulations for questions addressed here. In this paper, we focus on image simulation for partially coherent microscopy methods. In Sec. 2, we demonstrate the accuracy of the toolbox by comparing the experimental bright-field and dark-field images of the Siemens star pattern from the MBL/NNF test target [16]. The comparison also illustrates the utility of the toolbox by elucidating the sensitivity of the dark-field microscope to subresolution changes in periodic or isolated specimens. In Sec. 3, we describe image simulation algorithms and their computationally efficient implementation. In the same section, we develop a theory for estimating optical transmission of the specimen from a priori information about its ultrastructure that may be available from electron microscopy or X-ray diffraction. In Sec. 4, we use simulations to propose that the structural dynamics of a biological nanomachine, called axoneme, can be detected with a properly designed high-resolution dark-field microscope in a live specimen.

## 2. Evaluation of accuracy of microlith using MBL/NNF target

As illustrated by simulations on the microlith website, in cases where analytical expressions can be written, results produced with microlith are an excellent match to analytical results. Before discussing the details of implementation in the next section, we validate the simulation results obtained with microlith against experimental bright-field image of an amplitude Siemens star and dark-field images of a transparent, phase-only Siemens star from the MBL/NNF test target. In doing so, we gain insights into the sensitivity of dark-field microscopy to nanometer scale changes in the mass distribution of periodic or isolated specimens.

*2.1. Bright-field images of the amplitude target*

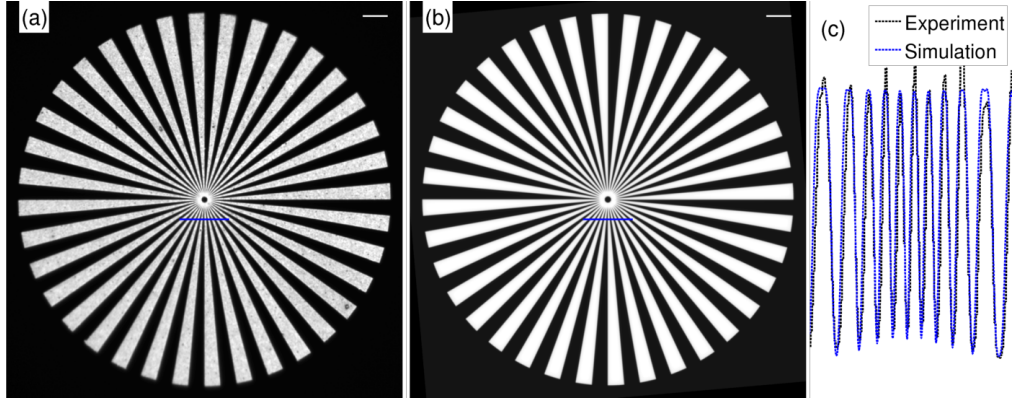

Fig 1. (a) Experimental and (b) simulated image of the MBL/NNF amplitude target using a bright field microscope. The scale-bar in top-right corner is 5um long. (c) intensity profiles of the experimental and simulated images taken along a 20 μm long line shown in blue in images (a) and (b). The simulated image faithfully reproduces the varying resolution and contrast found in the experimental image.

To check the accuracy of partially coherent images simulated using microlith, we first compare experimental and simulated bright-field images of the Siemens star pattern from the MBL/NNF amplitude target [16]. The star pattern was etched by electron beam lithography into a 40nm thick aluminum layer deposited onto a microscope cover glass. The 36 wedge pairs of the star pattern create an azimuthal square grating whose period $p$ increases with increasing radius $r$ from the center, specifically, $p = 2\pi r / 36$. The coverslip was bonded with a thin layer of Permount (refractive index 1.515, Fisher Scientific) to a microscope slide, with the aluminum layer facing the slide. The pattern was trans-illuminated with quasi-monochromatic light (halogen lamp with 577/10 nm interference band pass filter) and oil immersion condenser of the same numerical aperture as the 100x/1.3 NA Plan Achromat oil immersion objective. The experimental image, shown in Fig. 1(a), was captured by a sCMOS camera with 6.5 μm pixel size.

As expected, the experimental image shows intensity modulations at the period of wedge pairs (1 period/$10°$), but with varying contrast. The simulated image shown in Fig. 1(b) captures the pattern of intensity in the experimental image very well, except for the irregular mottle due to debris and irregularity in etching. Comparison of the experimental and simulated intensity profiles at a distance of 5 μm from the center illustrates that simulation captures the experimental modulations accurately. Near the center of both, the experimental and simulated image, the wedge pairs are no longer resolved as their periodicity becomes smaller than the resolution limit of the employed optics.

## 2.2. Dark-field images of the phase target

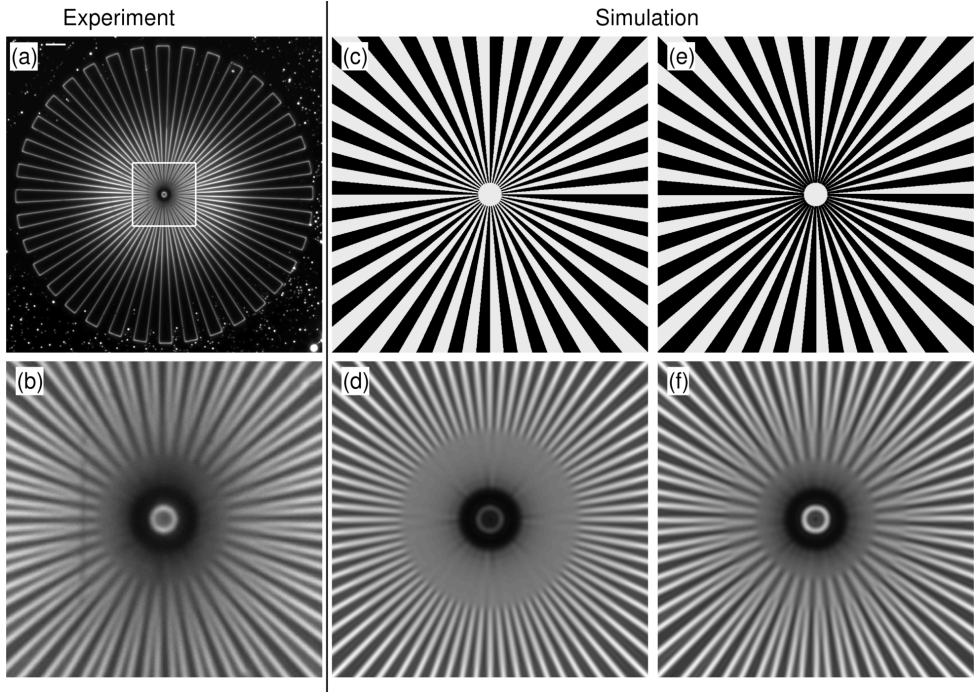

Figure 2 : Images of the Siemens star of the MBL/NNF phase target recorded with a dark-field microscope: (a) experimental image of the complete star pattern and (b) zoomed central region of (a). (c) zoomed center of a simulated pattern with equal sized wedges (black wedges represent etched silica layer, white wedges represent remaining silica layer). (d) simulated image of (c). (e,f) are images corresponding to (c,d), but assuming over etching by 4 nm. Scale bars: (a) 5µm, the central disk at the center in all images has a diameter of 1.2µm.

Figure 2 shows the experimental image of the Siemens star pattern of the MBL/NNF phase target recorded with a high-resolution dark-field microscope and images computed by microlith simulating the same optical setup. The phase target is etched into a 90nm thick fused silica layer (refractive index 1.46) deposited onto a microscope cover glass and etched using E-beam lithography [16]. As in the amplitude target, the wedge pairs create an azimuthal square phase grating. The cover glass with etched pattern is mounted on a microscope slide with a thin water layer between the pattern and the slide. The pattern was trans-illuminated with quasi-monochromatic light (halogen lamp with 546/40 nm band pass filter) and oil immersion condenser with a Phase 4 annulus in the front aperture. The average NA of the annulus was 1.15, with a lower limit of 1.1 and an upper limit of 1.2. The target was imaged with a 100x Plan Achromat oil immersion objective with aperture iris, which was set to 1.0 for rejecting the direct light and creating dark-field illumination. The experimental image was captured by the sCMOS camera with 6.5µm pixel size.

The experimental dark-field image in Fig. 2(a) shows the expected light scattering by the edges of the wedges, with the interior of every wedge nearly black, except near the center. We note that the interior of either type of wedge is black, the one with the silica standing and the other with the silica etched away. Fig. 2(b) shows the imaged pattern near the center, where we make the observation that, as we get closer to the center, the periodicity of the edges is replaced by the periodicity of the wedges, which is twice the edge periodicity. At even smaller radii, the dark-field microscope no longer resolves the wedges, which have periodicities that are smaller than the resolution limit of the employed optics.

In Fig. 2(c), we show a simulated phase pattern assuming perfect square etching, i.e. at a given radius the width of one type of wedge is exactly the same as the width of the other type of wedge. The image of such a target simulated with microlith is shown in Fig. 2(d). At first sight, it seems surprising that the simulated image does not match with the experimental image of Fig. 2(b) over a range of radii in which the edge periodicity is replaced by the wedge periodicity. We found the solution to this puzzle in assuming a slight over-etching of the etched wedges, which were then found to be wider by about 4 nm compared to the wedges in which the silica remained standing. Fig. 2(e) shows a simulated phase target that is over-etched by 4 nm, represented on a simulation grid with a sampling interval of 2nm, to adequately capture such fine detail. Fig. 2(f) shows the corresponding dark-field image simulated by microlith and demonstrating the exact same features as the experimental image. In appendix A1, we employ wave optical analysis to find out why the over-etching by 4nm indeed leads to the features observed in the experimental image.

The finding that the target could be over-etched by 4nm was subsequently confirmed by examining electron micrographs that were recorded when targets were manufactured. The analysis presented in Fig. 2 and appendix A1 led us to the discovery of sub-resolution deviations of a few nanometers from the intended design in the E-beam lithography process. This observation is a powerful illustration of (1) the sensitivity of the dark-field microscope to sub-resolution structural change in a periodic specimen and (2) the accuracy of our simulation code. As discussed in the context of imaging of axoneme in Sec. 4, high-resolution dark-field microscope can also be used for quantitative analysis of sub-resolution *changes* in the mass distribution of filamentous assemblies.

## 3. Partially coherent image simulation using microlith

### 3.1. Choice of optical model for partially coherent image simulation

Partially coherent imaging systems have been described in spatial, spatial frequency, and joint space-frequency (or phase-space) domains. The first model to account for the effect of an extended incoherent source on imaging is the 'sum over source' (SOS) model – proposed by August Köhler (see [17], pp.436), student and co-worker of Ernst Abbe. Köhler represented the total intensity due to an extended incoherent source as a superposition of intensities produced by mutually incoherent points of the source. Each point illuminates the specimen with an inclined plane-wave and the image due to the point is computed using Abbe's coherent diffraction theory.

Van Cittert and Zernike proposed the use of 'degree of coherence' of the source to account for the effect of an extended source [3]. In optical lithography, a partially coherent imaging system is commonly described in the spatial frequency domain using a transmission cross coefficient (TCC), originally proposed by Hopkins [18]. Hopkins's TCC is a 4D quantity for 2D images with the key benefit that the model separates the imaging system and the specimen. Such a separation is attractive for lithography simulations despite the need to store 4D quantities, because the goal is to evaluate images produced by a library of small features. To evaluate biological images generated by various microscopy methods, one needs to simulate larger patterns (such as the specimens evaluated in this paper) with variable microscope settings (such as defocus) for different contrast methods. When simulating a 3D image of a specimen, one needs to calculate the TCC for every focal plane, thus loosing the computational efficiency afforded by separation of the system and the specimen. Simulating a specimen area of $50 \times 50 \mu m^2$ with a pixel size of $200 \times 200 nm^2$ (typical pixel size needed for Nyquist sampling) requires a simulation grid of $250 \times 250$ pixels. To store the complex TCC coefficients as single-precision floating point numbers for simulating the $250 \times 250$ image, one needs $(250 \times 250 \times 8)^2$ bytes $= 250$ GB, which is a prohibitively large amount of memory.

Like Hopkins' TCC representation, phase-space representations of partially coherent imaging also provide the advantage of separating the imaging system and the specimen in their simulation models, in addition to providing new insights into the imaging process by representing the specimen and imaging system in a joint space-frequency domain [19]. However, phase-space models also require storage of 4D quantities and therefore large memory capacity. Due to the intractably large memory required to store 4D quantities for biological specimens, we have chosen to implement spatial-domain image simulation algorithms that need to store only 2D quantities. With the optimized algorithm implemented in microlith, we could perform parallel simulations of images of the Siemens star sampled over a $4001\times 4001$ grid under two contrast methods (dark-field and bright-field) with less than 1.5GB memory. While it is possible to reduce the required memory for 4D representations by breaking up the simulation space in smaller tiles, doing so brings up the problems of stitching the results.

One class of models aimed at reducing the computational and memory cost of partially coherent computation approximate the partially coherent imaging system as an incoherent sum of coherent imaging systems. Such methods typically rely on a matrix formulation of the partially coherent image computation [20]. Examples of such models are sum of coherent systems [21], optimal coherent approximation [22], coherent mode decomposition [23,24], and eigenanalysis of modified TCC matrices [25]. The essential idea behind such models is to find a smaller set of, preferably orthogonal, basis functions that span the vector space of images produced by the imaging system, when either the pupils of the imaging system or the properties of the specimen are subject to certain constrains. Such constrains frequently arise in optical lithography simulations, for example, when designing phase-shifting masks [22] that can assume only a small set of values, such as $\{0,1,-1\}$, or when the source and specimen can be described as a collection of small patches [25]. But, in biological microscopy simulations, it is difficult to place such constraints on the specimen or imaging system, because they typically employ illumination apertures that are the same size as imaging apertures, the imaging pupil can be complex valued due to optical processing (e.g., phase contrast and DIC) or aberrations, and specimens in general have asymmetric complex-valued spectra. Moreover, constructing the coherent decomposition of partially coherent systems requires applying singular value decomposition to large matrices, typically the TCC itself or its modified version [25], again posing potential memory bottlenecks. Note that in the sum over source (SOS) representation, each plane wave produced by an individual source point is orthogonal to other plane waves, each spatial frequency of the specimen is orthogonal to other spatial frequency, and each point in the imaging pupil is independent of any other point. Therefore, SOS representation provides a complete and non-redundant basis for representing the vector space of images produced by a general scalar partially coherent imaging system. While it may be beneficial to develop coherent mode approximations specific to popular image contrast methods, there is a need for a 'reference implementation' whose accuracy has been validated against experimental data. Therefore, we chose to implement a computationally optimized version of the SOS algorithm without coherent mode approximation and to check its accuracy against known analytical cases, experimentally established traits of partially coherent contrast methods, and images of well-characterized test objects.

*3.2. Computationally optimized implementation of sum over source algorithm*

With quasi-monochromatic partially coherent illumination, the 2D image recorded by a microscope, according to the sum over source algorithm, is:

$$I(x,y) = \iint S(\xi,\eta)\left|\mathcal{F}[T(f_x-\xi, f_y-\eta)P(f_x,f_y)]\right|^2 d\xi d\eta, \tag{1}$$

where, $I(x,y)$ is the image intensity, $S(\xi,\eta)$ is the source intensity distribution, $\mathcal{F}$ represents the 2D Fourier transform between the objective back focal plane (BFP) and the image plane, $T(f_x,f_y)$ is the spectrum of the object or specimen produced in the objective

BFP, and $P(f_x, f_y)$ is the amplitude of the imaging pupil. The above equation represents the fact that each source-point $S(\xi, \eta)$ illuminates the specimen with an inclined plane-wave that produces a shifted copy of the specimen spectrum $T(f_x, f_y)$ in the back focal plane of the objective, where the pupil function $P(f_x, f_y)$ acts as a filter.

While results are displayed in the real coordinates, the computations are carried out in normalized optical coordinates, where spatial dimensions are expressed in the units of $\lambda / NA_o$ and spatial-frequency dimensions are expressed in the unites of $NA_o / \lambda$, where $NA_o$ is the numerical aperture of the objective and $\lambda$ is the illumination wavelength. For efficient simulation of 3D images of thin specimens, microlith implements a version of the SOS algorithm that permits parallelized execution on multi-core computers and general-purpose graphics processing units (GPU). It can be written as follows:

$$I(x,y,z) = \sum_{i=1}^{N_s} S_i \left| \mathcal{F}[T(f_x - \xi_i, f_y - \eta_i) P(f_x, f_y, z)] \right|^2 \quad (2)$$

where, $i$ indexes only the non-zero pixels of the source distribution and $z$ is the amount of defocus. $S_i$ is the source intensity and $(\xi_i, \eta_i)$ are the shift in the specimen spectrum caused by the $i^{th}$ source pixel. The defocus-dependent imaging pupil $P(f_x, f_y, z)$ is computed only once for simulating multiple specimens, thereby providing a degree of separation between the specimen and system similar to spatial-frequency and phase-space models.

Above strategy makes it possible to transfer the vectors $S_i, \xi_i, \eta_i$, 2D specimen spectrum $T(f_x, f_y)$, and 2D defocused imaging pupil $P(f_x, f_y, z=a)$ to the GPU; compute intensity due to source points in parallel; and accumulate the resultant intensities in the matrix $I(x,y,z=a)$. This approach minimizes the memory transfers between main memory and GPU memory, which is typically the bottleneck when utilizing GPU. Table-1 summarizes the speed-up achieved due to above parallelization. The comparison is for simulation of 3D bright-field image at 11 focal depths, when computed on a single processor core running at 3.6GHz vs on a GPU running 448 cores at 1.15GHz .

**Table-1 Speed-up achieved by parallelized implementation of sum over source algorithm**

| Size of the simulated volume | Simulation time on one CPU core $t_{CPU}$ (s) | Simulation time on GPU $t_{GPU}$ (s) | Speed-up $t_{CPU} / t_{GPU}$ |
|---|---|---|---|
| 128x128x11 | 52 | 32 | 0.62 |
| 256x256x11 | 310 | 67 | 4.63 |
| 384x384x11 | 917 | 126 | 7.23 |
| 512x512x11 | 2040 | 192 | 10.62 |

As seen from above table, proposed parallelization quickly leads to speedup of an order of magnitude for moderate sized simulation volumes. The results presented in Fig. 5 are 2501x2501x11 in 36 minutes, whereas simulation of the same volume even using 12 CPU cores required approximately 7 hours. We are working towards speeding up the computation by another order of magnitude with an implementation that optimally utilizes multiple CPU cores and multiple GPUs.

*3.3. Radiometric consistency*

We devised important numerical optimizations that ensure that the simulated intensity is proportional to the image recorded by an ideal photo detector, a trait we name *radiometric*

*consistency*. To ensure radiometric consistency (over the chosen simulation grid) when either the specimen or the imaging system is varied, we normalize the computed intensity such that the image of a sub-resolution pixel (that approximates a point) is proportional to the area of the imaging pupil ($NA_o^2$), area of the illumination pupil ($NA_c^2$), and the area of the pixel. We choose normalizations such that the image of a point has the peak value of unity for $NA_o = 1, NA_c = 1$, and pixel size of 100nm. To ensure accurate simulation, the simulation grid needs to extend beyond any feature that transmits light by at least the size of the point spread function of the imaging path. When above simulation conditions are met, the simulated integrated intensities are found to follow the traits of physically recorded intensities – e.g., the integrated intensity is invariant with respect to defocus in the absence of confocal gating in the detection path.

*3.4. Imaging with polychromatic illumination*

While the experiments and simulations reported in this paper are carried out with quasi-monochromatic illumination, polychromatic illumination is frequently employed for microscopy and macroscopy. To enable image simulation with polychromatic illumination, microlith allows the user to specify the spectral distribution of the light source, spectral sensitivity of the camera, specimen dispersion, and chromatic aberration of imaging path. Chromatic aberration can be classified in two categories, lateral and longitudinal, both arising from the change in effective focal length of the objective and/or tube lens due to dispersion in the glass used for their manufacture [26].

A practical method of assessing lateral chromatic aberration is to image multi-color sub-resolution beads and measure the relative displacement (chromatic shift) and relative scaling (chromatic scaling) from the recorded point spread functions. microlith allows the user to input the lateral chromatic shift and lateral chromatic scaling at each wavelength. Even in the absence of chromatic aberration, the image at one wavelength is scaled version of the image at another wavelength, purely because of diffraction - the scaling factor being the ratio of two wavelengths. We specify lateral chromatic scaling as a factor beyond the scaling due to diffraction.

In the presence of lateral chromatic aberration and negligible dispersion by the specimen, the image due to polychromatic illumination can be computed simply as shifted (due to chromatic shift), scaled (due to wavelength and chromatic scaling), and weighted (due to spectral distribution of the source and spectral sensitivity of camera) summation of the image computed at the center wavelength. This approach does not incur significant computational cost and is reasonably accurate for imaging non-dispersive biological specimens using well-corrected imaging systems.

However, if the longitudinal chromatic aberration is significant or if the specimen has significant dispersion, it is essential to compute the imaging pupil or specimen transmission separately at each wavelength. While the memory requirements for computing polychromatic image using sum over source algorithm will not be onerous, the computation time will increase by the number of wavelengths at which the image is calculated, irrespective of which model is used for image computation. The optimal sampling interval along the wavelength depends on the dispersion by the specimen and the dispersion by the optics. Accounting for significant dispersion by the specimen or imaging system in a computationally efficient manner is an area of future work and will be facilitated by current efforts directed at developing sum over source algorithm optimized for multiple CPU-cores and multiple-GPUs.

*3.5. Estimating specimen transmission from its ultrastructure and coherence volume*

The first step in simulating the image of a thin specimen is to estimate its transmission (consisting of absorption and optical path difference) profile. We present a first-order theory to compute transmission of a thin specimen from its ultrastructure that may be available from electron microscopy or X-ray diffraction, as is the case for the axoneme discussed in Sec. 4.

The strength and angular distribution of the light scattered by a specimen (e.g., protein assembly) embedded in a medium (e.g., water) depends on the distribution of the dielectric polarizability in excess of the surrounding medium. To estimate the effective refractive index from the ultrastructural information, we combine two theories: (a) the theory of an effective dielectric constant of a coherently illuminated dielectric mixture, originally proposed by Weiner and used by us to estimate the form birefringence of single microtubules and axonemes [27], and (b) McCutchen's theory that the 3D mutual coherence is the 3D Fourier transform of the source intensity distribution mapped onto the Ewald sphere [28] analogous to a generalized imaging aperture [29]. In the following, we calculate the *effective optical path difference due to structure of a thin specimen and partially coherent illumination,* which is then imaged according to the point spread function or the imaging pupil.

We define the *coherence volume* as the coherently illuminated volume around *every point* $(x,y)$ of the specimen. The extent and shape of the coherence volume is related to the illumination aperture. According to the Van Cittert–Zernike theorem [3], the mutual coherence in the specimen plane (a function only of the distance between two points) is the Fourier transform of the intensity distribution of the source. McCutchen [28] extended this idea to 3D mutual coherence. Note that the coherence volume is conceptually different from the extent of the point spread function of the imaging path, although numerical relationship between the illumination intensity and the 3D mutual coherence is the same as the numerical relationship between the imaging pupil's amplitude and the 3D amplitude point spread function. We have implemented the computation of the 3D mutual coherence in microlith based on this theory. We specify the coherence volume as a volume over which mutual coherence is at least 90%.

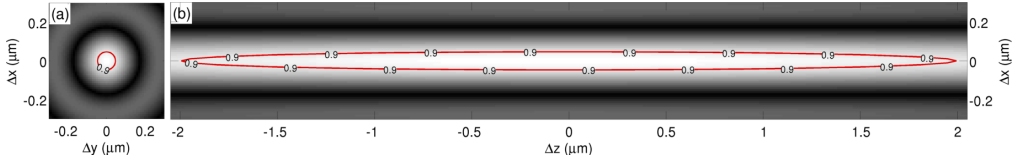

Figure 3: Mutual coherence due to dark-field source of NA 1.1-1.2 with center wavelength 0.546µm : (a) In the specimen plηane, (b) along the optical axis. 3D coherence is function of distance $(\Delta x, \Delta y, \Delta z)$ from point $(x, y)$ on the specimen. Red contour shows the distances from any point in the specimen within which the coherence is greater than 90%.

Figure 3 shows the 3D mutual coherence simulated for dark-field illumination that was used for recording the images shown in Fig. 2. As expected for illumination with a ring-shaped illumination, the coherence volume in Fig. 3 is a highly elongated prolate, with a radius of 50nm in the specimen plane and an axial extent ($Z_c$) of 2µm.

Over a coherently illuminated volume, a thin biological specimen introduces electric polarizability proportional to the fraction of the volume occupied by its protein mass. For simplicity, let us assume that the ultrastructural components of the specimen have the same dielectric permittivity denoted by $\epsilon_p$. Also assume that the structure is embedded in a medium with dielectric permittivity $\epsilon_w$. At position $(x,y,z)$, the protein occupies a certain fraction, $f(x,y,z)$, of the coherence volume. Once the size and shape of the coherence volume is computed using microlith, $f(x,y,z)$ is computed from the discretized ultrastructure of the specimen by placing the coherence volume at each voxel and counting the fraction of voxels occupied by the specimen out of the total number of voxels within the coherence volume. Thus, $f(x,y,z)$ is essentially the convolution of the specimen ultrastructure with coherence volume normalized by the size of the coherence volume. When the mixture is illuminated coherently, the effective dielectric constant is [31],

$$\epsilon(x,y,z) = \epsilon_w + f(x,y,z)(\epsilon_p - \epsilon_w), \tag{3}$$

where, $\epsilon_p - \epsilon_w$ indicates the polarizability introduced by the biological structure in excess of the embedding medium. The distribution of effective refractive index is given by $n(x,y,z) = \sqrt{\epsilon(x,y,z)}$. Therefore, the optical path difference profile is

$$OPD(x,y) = \int_{Z_c} \frac{2\pi}{\lambda}[n(x,y,z) - n_w]dz = \int_{Z_c} \frac{2\pi}{\lambda}[\sqrt{\epsilon(x,y,z)} - \sqrt{\epsilon_w}]dz \tag{4}$$

where, $Z_c$ is the axial extent of the coherence volume.

The specimen's transmission function is given by,

$$t(x,y) = \exp[iOPD(x,y)]. \tag{5}$$

The above discussion can be extended in a straightforward manner to estimation of absorption profile of a thin specimen by assuming that dielectric permittivity has an imaginary component. The extension of above theory to imaging of thick (relative to coherence volume) specimens is an area of future work.

### 4. Quantification of structural changes in axoneme using dark-field microscope

In this section, we discuss a biological application of dark-field microscopy facilitated by simulations using microlith. The axoneme [Greek: *axon* axis + *nema* thread] forms at the same time the 'skeleton' and the 'motor' of flagella and cilia in eukaryotic cells. It is a cytoskeletal nanomachine with diameter $\approx 200$ nm and is usually many microns long (see Fig. 4). In most species, it consists of microtubules arranged in a highly conserved 9+2 structure of 9 microtubule doublets (fused microtubule singlets) forming a ring around 2 microtubule singlets at the center [32,33].

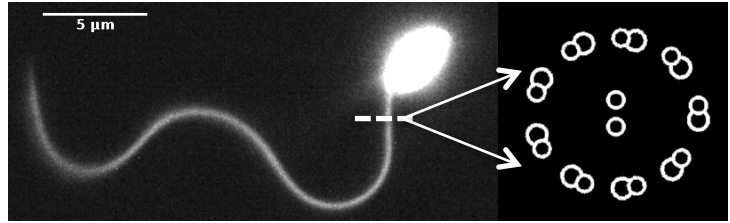

Figure 4: Dark-field image of a sea urchin sperm and arrangement of microtubules in its cross-section, which are key components of axoneme. Diameter of axoneme is ~ 200nm. The image was acquired with sCMOS camera using the same configuration as in Fig. 2.

Dynein motor proteins link the outer doublets together and effect the bending of the entire assembly. In contrast to the static composition of the axoneme, the motile mechanisms of the molecular constituents that lead to the periodic beat pattern of flagella and cilia remain a mystery to this day, despite great efforts by a large number of scientists. The difficulty arises in part from the fact that the axoneme is a complex molecular assembly of relatively small diameter that does not lend itself to dynamic X-ray analysis, like muscle fibers do [34]. Therefore, we cannot obtain high-resolution structural data at length and time scales in live axonemes that could reveal the molecular dynamics giving rise to the periodic beating of this ubiquitous organelle.

However, the sensitivity of a high-resolution dark-field microscope configuration to structural changes at the length-scale of 10nm (discussed in Sec. 2.2) makes dark-field a promising method for probing dynamic structural changes in sparse structures such as a beating axoneme. In appendix A2, we clarify the sensitivity of the dark-field microscope to sub-resolution changes in the mass of filamentous structures by comparing the experimental and simulated dark-field images of line features available in the MBL/NNF phase target.

Based on the published cryo-EM measurements [33] of the structure of sea urchin axonemes and personal communications with Dr. Daniela Nicastro, we judged that the main contributions to the polarizability by the axonemal assembly come from the microtubules, which are the only continuous structures along the length of the axoneme. Therefore, we choose to simulate the axoneme as a 9+2 structure of microtubules. In support of this assumption, we note that the same assumption has led to an accurate estimate of the birefringence of sea urchin axonemes [27]. EM measurements on bent Chlamydomonas axonemes [35] have shown that the distance between microtubule doublets increased in the plane of the bend. Changes in cross-sectional distribution of microtubules are also expected based on atomic force microscopy measurements [36]. We also note that all experimental evidence available today indicates that the axoneme beat is not associated with large-scale (>200 nm) redistributions of mass along the length of the axoneme. Instead, the structural changes that seem to occur in the axoneme during the beat can be described by a change in its effective diameter. As our simulation demonstrates, even small changes in effective diameter can be measured by dark-field microscopy.

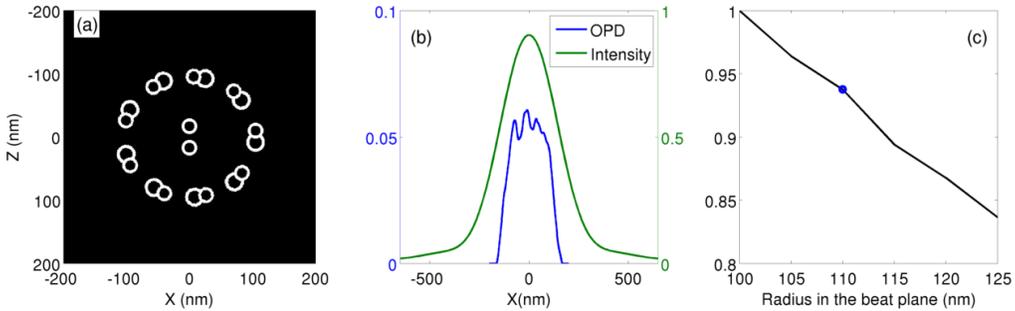

Figure 5 Simulated dark-field intensity as a function of change in the axoneme cross-section: (a) pixel-map of singlet and doublet microtubules, black represents solvent, white tubulin mass. (b) optical path difference $OPD(x)$ in radians and simulated dark-field intensity profile for a microscope configuration specified in Fig. 2. (c) Normalized mean intensity as a function of radius of the axoneme cross-section in the beat plane. Different frames of the movie show above plots as the axoneme's radius in beat plane increases, but total area inscribed by doublet microtubules is preserved.

Next, we simulate the effect of change in axoneme radius in beat plane (while the area inscribed by doublet microtubules remains constant) on darkfield image recorded by the microscope described in Sec. 2.2. Based on cryo-EM data (see Fig. 1(A) and Fig. 4(D) in [33]), we modeled the radii and positions of singlet and doublet microtubules to produce a pixel-map of microtubules in the cross-section of an axoneme. Figure 5(a) shows the cross-section of axoneme modeled as 9+2 structure of microtubules. The cross-section is assumed to lie in the XZ plane. The white pixels in Fig. 5(a) reflect the cross-section occupied by tubulin protein, whereas black pixels reflect the solvent (water). Since, the diameter of the axoneme is ~200nm, it is sufficient to simulate the cross-section of the axoneme over 400nm x 400nm grid with sampling of 1nm.

As indicated earlier, we consider the mass per unit length of the axoneme to be constant along its axis. For estimating the volume-fraction $f(x,z)$ of protein mass as a function of position in the direction perpendicular to the axoneme's long axis, we convolve the axoneme cross-section with the cross-section of coherence volume delineated by red line in Fig. 3(b). Subsequently, eqs. 3 and 4 are used to calculate $OPD(x)$ by assuming the permittivity of tubulin $\epsilon_p = 1.512^2$ as measured by Sato, Ellis, and Inoué [30], permittivity of water $\epsilon_w = 1.33^2$, and the axial extent of the coherence volume $Z_c$ equal to 2000nm.

Figure 5 simulates the axoneme cross-section deformed by different amounts in the beat-plane and the corresponding dark-field intensities. Figure 5(a) shows the cross-section, Fig. 5(b) shows corresponding profiles of $OPD(x)$ and simulated dark-field intensity profile, normalized such that the peak intensity across all simulated cross-sections is 1. As seen in Fig. 5, the increase in axoneme's radius causes broadening of the OPD and reduction in the peak intensity recorded by the dark-field microscope. Since the total protein mass in the cross-section of the axoneme does not change, the total protein fraction or the area under $f(x)$ remains constant, which in turn implies that the area under $OPD(x)$ remains constant. *Therefore, any change in recorded dark-field intensity implies a change in the distribution of mass over the cross section of the axoneme.* As seen from the animation of simulated intensities in Fig. 5(b), the peak value of the dark-field intensity drops monotonically as the axoneme expands.

We integrate the intensity perpendicular to the axoneme's length (along its cross-section) to improve the signal to noise ratio. Like the peak intensity, the integrated intensity increases with tighter packing of the microtubules in the axoneme. However, unlike the peak intensity, the mean intensity does not change with the focus of the axoneme. In wide-field imaging, the intensity integrated in the focal plane is independent of defocus in the absence of depth gating (e.g., depth gating by a confocal pinhole). Figure 5(c) shows the integrated or mean intensity normalized to its maximum value. The analysis shows that the mean intensity recorded by the dark-field microscope varies inversely with the 'tightness' of microtubule packing in the axoneme.

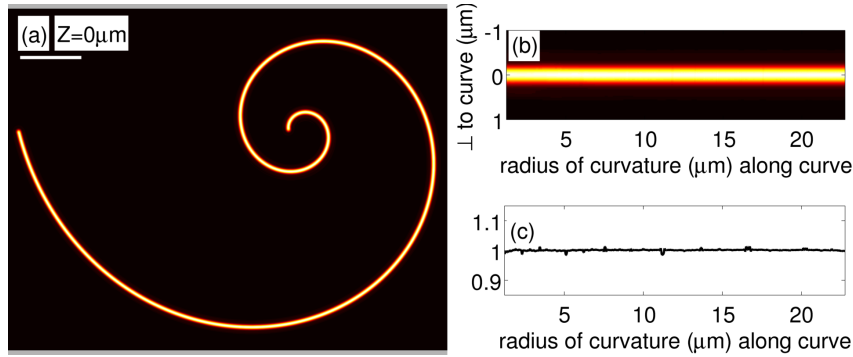

Figure 6: (a) simulated dark-field image of a cylinder bent in the shape of logarithmic spiral with gradually increasing radius of curvature (scale bar: 5 um), (b) intensity distribution perpendicular to the cylinder as a function of radius of curvature, and (c) mean intensity as a function of radius of curvature. The images are simulated assuming the same optical setup as in Fig. 2.

During experiments, recorded intensity may be affected by sharp bends along the length of the axoneme and parts of the axoneme that may go out of focus while beating. The axoneme is a relatively stiff organelle with radius of curvature of at least a few microns. While some flagella have 3D helical beat pattern, the model system we use (sea urchin sperm) has a planar beat pattern. To test how the shape along the long-axis of the axoneme and the defocus affect interpretation of mean intensity, we simulated 3D dark-field image of a uniform cylinder of refractive index 1.512 and diameter 200nm bent such that it presents a large range of radii of curvature to the imaging system. For the shape of the axoneme, we chose a logarithmic spiral [38], described by the equation $r = e^{\theta/4}$, in polar coordinates $(r,\theta)$. At any point on the spiral, the radius of curvature is simply $r$. The analysis of its intensity profile are shown in

Fig. 6. Figure 6(a) shows the simulated image and 6(b) shows the intensity profile normal to the local tangent as a function of curvature of the spiral. Figure 6(c) shows that the mean intensity is almost constant for radii of curvature down to a micron barring occasional noise.

For a conceptual understanding of our simulation results, we note that the axoneme is a long, but thin object whose diameter is close to the resolution of the microscope. When the diameter of such rod-like structure increases, the light it scatters into the plane perpendicular to the long axis is more and more forward directed, i.e. close to the direction of the incoming light [37]. As more of the scattered light is forward directed, it is lost into the range of angles not collected by the objective lens in the dark-field setup. Therefore, the brightness of the axoneme image decreases as the diameter of the axoneme increases. Note that above simulations are based on the diameter distortions measured in fixed axonemes. The change in diameter in a live and beating axoneme may be different, leading to variation in integrated intensity differnt than 8% predicted by the simulation in Fig. 5. We have established the measurability of changes in integrated intensity of a beating axoneme, which will be reported and analyzed in a manuscript under preparation. Thus, the intensity in dark-field images of the axoneme reports not only its shape, but also changes in the tightness of microtubule packing correlated with the shape.

## 5. Conclusion

With the experimental results and theoretical analysis presented in this paper, we tested the accuracy of partially coherent image simulations with microlith. We recorded experimental bright-field and dark-field images of a Siemens star etched into a thin aluminum or $SiO_2$ film. Comparing experimental images with simulations demonstrated the accuracy and versatility of the microlith code and highlighted the sensitivity of dark-field imaging to nm variations in pattern layout.

We described the algorithmic choices made for memory efficient and computationally efficient simulation of radiometrically accurate images. We also developed a connection between Wiener's theory of effective dielectric constant of coherently illuminated mixture and McCutchen's theory of relationship between 3D coherence and generalized source. This connection allows us to estimate effective transmission of a thin specimen illuminated by partially coherent light. This approach can be extended to compute effective 3D refractive index distribution from the properties of a thick object and partially coherent illumination, which is a promising direction for addressing the challenging problem of modeling 3D partially coherent image formation.

We then used microlith to simulate how changes in the cross-sectional distribution of microtubules in the axoneme affect the intensity recorded by a high-resolution dark-field microscope. We developed an optical model for the axoneme which leads us to hypothesize that nanometer scale changes in the positions of the microtubules in the axoneme cross-section can be detected by dark-field microscopy. Our simulations suggest that monitoring the changes in mean intensity of dark-field images of a functional, beating axoneme can provide a fast and accurate measure of changes in the distribution of density of microtubules in the cross-section of the axoneme. In an article to be published, we report measurements of the dark-field mean intensity of beating axonemes as a function of distance to the axoneme base and time. We then relate the simulation results reported here to a model of structural changes that occur in synchrony with the beat pattern.

The algorithms in microlith have been used to analyze performance of coherent [14] and incoherent imaging systems [15]. Thus, microlith can serve as a general image simulation tool for biological microscopy methods. Apart from gaining insights about the image formation and informing the analysis of partially coherent data as described in this paper, a promising use of microlith is to compare the transfer properties of quantitative phase-retrieval methods in terms of the spatial frequencies of the specimen. Phase-retrieval [or retrieval of specimen anisotropy requires acquisition of multiple images with systematically varied pupils and algorithmic recovery. The entire process can be characterized by simulating the chosen model specimen with appropriate pupils, and by evaluating the contrast transfer from the phase image produced by the retrieval algorithm. In the future, we intend to extend microlith to incorporate simulation of polarized light microscopy and imaging of 3D specimens. We hope that the accuracy and the open source character of microlith toolbox will facilitate quantitative use of partially coherent systems for biological imaging.

**Appendices**

*A1. Wave optical analysis of dark-field image and simulations shown in Fig. 2*

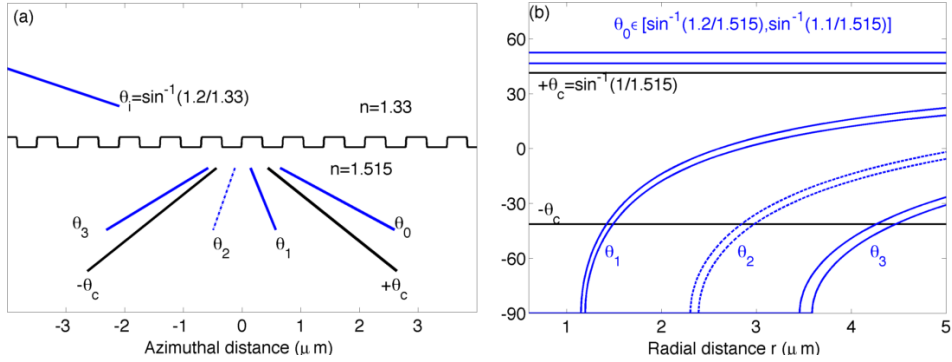

Figure 7: Analysis of intensity pattern observed in Fig. 2: (a) Illustration (drawn to scale) shows the angles of diffraction orders generated by the periodic phase grating at the radius of 4μm due to the steepest illuminating ray. (b) shows the range of angles occupied by the diffraction orders generated by the periodic grating at radii (r) ranging from 0.6 − 5μm. Note that $\sin\theta_N = \sin\theta_0 - N\lambda(36/2\pi r)$, where, $N = 1,2,3$.

Beyond the radius of 6μm, the experimental image in Fig. 2(b) matches very well with simulated images in Figs. 2(d) and 2(f). Going towards the center in Fig. 2(b), the azimuthal period of intensity changes from a period of edges (2 periods/10°) to a period of wedge pairs (1 period/10°) at the radius of 4.8μm. However, the simulation in Fig. 2(d) assuming an exact square grating (36 symmetric wedge pairs) shows that the contrast vanishes at the radius of 4.8μm. The experimental image shows intensity modulation of 1 period/10° up to the radius of 3μm, uniform intensity up to the radius of 1.5μm, and finally almost no intensity up to the bright ring at radius of 0.6μm. Simulation in Fig. 2(d) on the other hand shows uniform intensity over radii of 4.8-1.5μm, dark-band over 1.5-0.6μm similar to experimental image, but much dimmer central bright ring. Thinking of the azimuthal phase profile as a linear grating as shown in Fig. 7(a), above result can be understood as follows: Since the target is immersed in water (refractive index 1.33), a hollow cone of light with an angular span $\theta_i \in [\sin^{-1}(1.2/1.33), \sin^{-1}(1.1/1.33)]$ illuminates the target. In region without grating structure, the light exits the coverslip with an angular span $\theta_0 \in [\sin^{-1}(1.2/1.515), \sin^{-1}(1.1/1.515)]$, which is also the angular span occupied by the undiffracted light (zeorth order). The wavelength of diffracted light produced in the coverslip is given by $\lambda = 0.546/1.515 = 0.36\mu m$. From the grating equation, the angles $\theta_N$ in which the diffraction orders generated by a periodic phase grating can propagate are given by $\sin\theta_N = \sin\theta_0 - N\lambda/d$, where $d = 2\pi r/36$ is the grating period at radius $r$ and $N$ the diffraction order. Using above equations, Fig. 7(b) plots the range of angles occupied by $0^{th}$ to $3^{rd}$ diffraction orders $(\theta_0, \theta_1, \theta_2, \theta_3)$ as a function of radius $r$. Since the imaging NA is 1, the steepest angle of the ray that can be collected after the coverslip (RI 1.515) is

$\theta_c = \pm \sin^{-1}(1/1.515)$, shown by black lines in Figs. 7(a and b). The interference of 1st and 2nd order gives rise to intensity modulation with the period of the square wave (i.e., period of the wedge pairs). The interference of 1st and 3rd order gives rise to intensity modulation with the period of the edges. Fig. 7(a) illustrates that at the radius of 4µm, only the 1st and 2nd orders can be collected. From Fig. 7(b), we see that the 1st, 2nd, and the 3rd orders enter the collection aperture at radii of 1.5µm, 3µm, and 4.5µm, respectively. Therefore, above analysis accurately explains the experimentally observed transitions in the intensity modulation as a function of diffraction orders produced by the azimuthal grating at different radii. *However, if the azimuthal pattern is an exact square wave, the diffraction spectrum consists of only odd numbered orders.* The 2nd order propagating at $\theta_2$ (shown by a dashed line) is not generated by an exact square grating, leaving only the 1st order at $\theta_1$ up to the radius of 4.5µm, which by itself cannot give rise to intensity modulation. Thus, above analysis and comparison of the experimental image (Fig. 2(b)) and simulated images (Figs. 2(d and f)) lead us to conclude that the etched pattern has a slight asymmetry in the azimuthal square wave that produces a $2^{nd}$ diffraction order and an intensity modulation in the image at the period of wedges. As shown in Fig. 2(f), we found that over-etching by 4nm provides the best fit between experimental and simulated data; including the bright ring of radius 0.6µm, which is due to a fully-etched ring around the central intact silica disk.

*A2. Sensitivity of dark-field microscope to subresolution change in structure of line features*

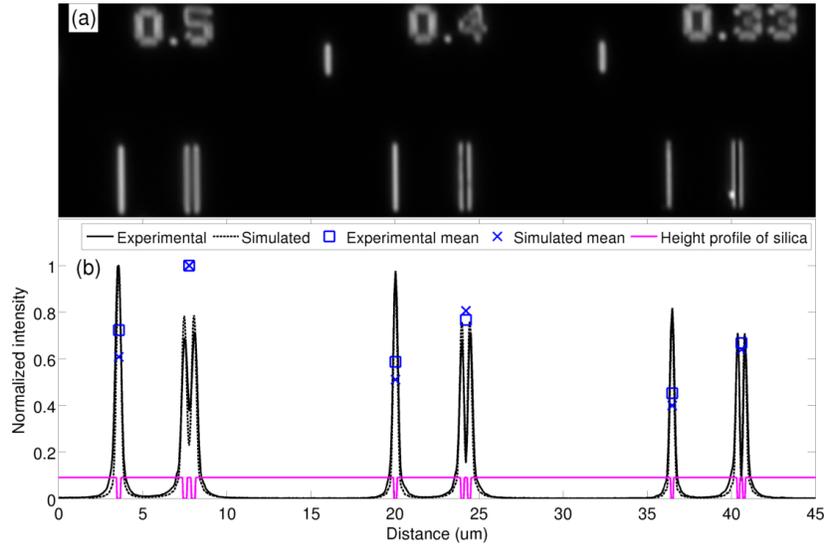

Figure 8: Single and double line features from MBL/NNF phase target imaged with dark-field microscope (a) experimental image, (b) simulated thickness profile (magenta) and simulated & experimental intensities. Markers show the total scattered intensity around line features. The number etched in (a) on top of each feature is twice the line width or line separation in microns.

In the MBL/NNF target, we chose line features of widths 250nm, 200nm, and 165nm shown in Fig. 8(a). The double lines are separated by the same distance as the width. Since the phase-target was immersed in water (R.I. 1.33), line features had an optical path difference of $(2\pi/0.577) \times (1.33 - 1.46) \times 0.09 = -0.04\pi$. The magenta line in Fig. 8(b) shows the simulated phase-profile for chosen line features. Figure 8(b) compares the experimental and simulated intensity profiles normalized by respective peak intensities and demonstrate excellent agreement between experiment and simulation. The markers in Fig. 8(b) show integrated intensity around the center of each single or double line feature for both experimental and simulated intensities. As detailed in Sec. 4, integrated or mean intensity is a useful and robust measure of the density distribution of isolated scatterers.

## Acknowledgements

This study was funded by National Institutes of Health grant RO1 EB002583. SM acknowledges postdoctoral fellowship from Human Frontier Science Program (HFSP) and useful discussions with Colin Sheppard, Naoki Noda, Tomomi Tani, Daniela Nicastro, Daniel Chen, Jerome Mertz, and Jean-Charles Baritaux.

## References and Links